\documentclass[fleqn]{article}

\title{Architectural Adequacy and Evolutionary Adequacy as Characteristics  of a Candidate  Informational Money}
\usepackage{stmaryrd,amssymb,amsmath,amsthm,url}

\theoremstyle{definition}

\newcommand{\FR}{\mathsf{FR}}

\author{\Large
	Jan A. Bergstra\\~\\
  {
	  Informatics Institute, Faculty of Sciences,
	  University of Amsterdam\footnote{%
	{Email: \texttt{j.a.bergstra@uva.nl.}}}
	}
}

\begin{document}
\maketitle

\begin{abstract}
For money-like informational commodities the notions of architectural adequacy and evolutionary adequacy 
are proposed as the first two stages of a moneyness maturity hierarchy. 
Then three classes of informational commodities are distinguished: exclusively informational commodities, strictly informational commodities, and ownable informational commodities. For each class money-like instances of that commodity class, as well as monies of that class  may exist. 

 With the help of these classifications and making use of previous 
assessments of Bitcoin, it is argued that at this stage Bitcoin is unlikely ever to evolve into a money. Assessing the evolutionary adequacy of Bitcoin is perceived in terms of a search through its design hull for superior design alternatives. 

An extensive comparison is made between the search for superior design alternatives to Bitcoin and the search for design alternatives to a specific and unconventional view on the definition of fractions. 
\\[5mm]
\end{abstract}
\tableofcontents
\section{Introduction}
In~\cite{BergstraL2013} and~\cite{BergstraL2013b} it has been argued that the phrase informational 
money may be helpful for understanding
moneyness issues about systems like Bitcoin as first described in~\cite{Nakamoto2008}. However, the notorious question whether or not Bitcoin constitutes a money is certainly not solved by simply asserting that it constitutes an informational money rather than a conventional money. Implicitly introducing more flexibility for the concept of informational money than one would allow for the notion of money will not solve conceptual and definitional problems.

In~\cite{BergstraW2014} and subsequently in~\cite{Swanson2014} it has been argued that Bitcoin provides us with a money-like informational commodity (MLIC), while in~\cite{Bergstra2014b} it has been argued that Bitcoin does not constitute a currency-like informational commodity with the understanding that  currencies must necessarily  be moneys  while some moneys may not be currencies.
Both papers provide criteria under which a money-like informational commodity may evolve into an informational money or an informational currency. That evolution is reversible in both cases. In other words,   an informational commodity may have an alternating status regarding moneyness and currency-ness. It is conceivable that, during a phase of uninterrupted  moneyness, an informational commodity switches back and forward with respect to its currency status.

\subsection{Objectives of the paper}
The first objective of this paper is to develop a methodology for making further progress on questions of  moneyness for Bitcoin and for similar systems. I will describe  a hierarchy of maturity levels for MLIC's 
and in particular I will suggest two new stages for that hierarchy: architectural adequacy and evolutionary adequacy. Below I will analyse in some depth how these levels may support the description of an effective and attractive research program concerning the plausibility of future moneyness of Bitcoin.

A second objective is to formulate general criteria for the classification of informational monies and MLIC's.
More specifically a classification in three forms of informational commodities is proposed from which further classification of MLIC's and of informational monies can be inherited or developed.

A third objective of the work reported in this paper is to develop a comparison between two lines of investigation: 
investigating the evolutionary adequacy of a specific candidate informational money, and investigating the 
long term sustainability of some specific views on the concept of a fraction. In spite of the vast difference between informational monies and fractions I claim that there is ample room for synergy between both lines of investigation.

\subsection{The Bitcoin design hull}
I hold that it is impossible to be able to assert categorical properties (like moneyness)  for Bitcoin without 
being able to assert the 
same or quite similar properties for variations of its design. Judgements about moneyness and currency-ness
require a general framework on which such judgements can be based and that framework is far more general than what can be obtained by considering a single money-like informational commodity.

This implies that a categorical discussion about the moneyness of Bitcoin necessarily must employ an abstract, that is architectural, view of Bitcoin technology. When claiming that Bitcoin provides a money-like informational commodity the categorical classification as such cannot be applicable to Bitcoin only. Rather than a judgement about Bitcoin that verdict will constitute a judgement about all systems in the design hull of Bitcoin.

With the {\em design hull of Bitcoin} I propose to indicate, in a fairly informal and open ended manner, all existing and forthcoming designs of money-like informational commodities that are similar in some sense to Bitcoin, either in terns of design, or in terms of objective or both. The situation is even more complex that this: some systems 
or system designs may be similar to Bitcoin and may be viewed as carriers of an informational commodity but may not be considered money-like. Therefore the design hull of Bitcoin extends beyond MLIC's
\subsection{Assessing Bitcoin as an instance of MLIC assessment}
Assuming that one agrees with the classification of Bitcoin as a money-like informational commodity than unavoidably one must accept that many alternative designs within the design hull of Bitcoin also produce (or would produce, when thinking hypothetically) a money-like informational commodity. Some judgements about the status of Bitcoin cannot be made without having developed a capability to produce similar judgements about MLIC's in general, or if that is too general about Bitcoin-like MLIC's.

An observation made in~\cite{BergstraW2014} is that given the conceptual/categorical classification of Bitcoin as 
a money-like informational commodity the question whether at some stage Bitcoin actually is a money can take all technicalities and all conceivable specifics into account. Even without any modification in architecture, design,
or implementation Bitcoin may move from being a non-money (negative moneyness) to being a money (positive moneyness). 

An MLIC may also be referred to as a candidate informational money, or simply as a candidate money in case 
the requirement that the underlying commodity is informational transpires from the context.

\section{Status assessment methodology for candidate monies}
Assuming that the assessment has been made that an informational commodity is money-like four further stages can be achieved. Each of these is subjective in the sense that different observers making use of different theories of money and different views on what constitutes a currency may come to different conclusions.

The conclusions of~\cite{BergstraW2014}  include the following observations: (i) Bitcoin has positive potential moneyness, 
(ii) the actual moneyness may fluctuate in time, and (iii) different observers may agree on potential moneyness of a system while disagreeing on precise criteria for actual moneyness. 

The  approach of~\cite{BergstraW2014}  to status assessment for Bitcoin, however, is too simple because although one may consider Bitcoin to be money-like, one may also hold the viewpoint that its architecture is flawed to the extent that no matter how its usage characteristics will develop in practice it will not qualify as a money. Below I will try to be more specific about the role of this aspect of assessment of moneyness of Bitcoin,
and I will phrase the discussion in terms of what I propose to call the architectural adequacy for moneyness of an MLIC.

Before giving any assessment of moneyness depending on acceptance, average size of transactions, spread of circulation, and intensity of usage, 
an assessment of its architectural adequacy needs to be made. In order to provide work on MLIC status assessments with some logical structure I propose to consider the following four stages or maturity levels  that a candidate money may (or may not) reach.

\subsection{Maturity levels for an MLIC}
The four levels mentioned below are not to be understood as discrete levels in a strict sense. In fact each 
level brings implicitly with it a degree of maturity rather than a firmly determined stage of maturity. 
\begin{description}
\item [Architectural adequacy (for moneyness):] The candidate money features a number of properties which together imply 
that in principle it can be classified as a money if state dependent (that is non-architectural) conditions are supportive of that.
 
For instance one may assume that the Euro system provides an architecturally adequate (though not informational) money. Even if its circulation decreases drastically and parallel monies emerge in most countries 
or even in all countries of the Eurozone it's not the architecture but a complex of contingent conditions that may cause a negative assessment about its moneyness. 

Architectural adequacy is static though subjective. Only by changing one's theory of money, or one's approach to applying that theory, the adequacy judgement for a specific candidate money may be changed. A positive judgement on architectural adequacy needs to incorporate a detailed view on what kind of money, that is
what subset of the functional moneyness criteria as mentioned in~\cite{BergstraL2013} is supposed to be 
met,  an MLIC might possibly constitute.

In~\cite{BergstraL2013} moneyness has been portrayed as a multidimensional notion. Each money can be thought of as a member of a space of possible monies. Monies may for instance 
differ greatly or subtly concerning matters of anonymity and privacy. 
\item [Evolutionary adequacy (for moneyness):] An architecturally adequate candidate money is evolutionary adequate if a development path towards moneyness can be imagined which is likely not to be blocked by
competing monies or by candidate moneys striving for moneyness status as well. Evolutionary adequacy is trivially positive 
once positive moneyness has been achieved. Evolutionary adequacy may be lost if a new design for a
candidate money appears, either on paper or in practice. Once lost, it is very unlikely for evolutionary adequacy to be regained. That may happen, however, if one or more competing informational moneys or candidate monies 
turn out to be unsustainable to the extent of being withdrawn from the market of (informational) monies so that the competitive pressure for the candidate money under assessment for evolutionary adequacy may be judged more positively after all.
\item [Moneyness:] In~\cite{BergstraW2014} it has been outlined under what conditions a candidate money may be classified as a money. These conditions are contingent, that is depending on the quantification of the actual state of the underlying system in terms of circulation and in terms of its share of transactions in comparison with transactions for other monies or candidate monies. Moneyness is not black and white, rather degrees of moneyness may be distinguished. 

By definition an architecturally inadequate candidate money can't develop into a money. For an architecturally adequate candidate money one may require that evolutionary adequacy coincides with plausibility of potential moneyness. If potential moneyness is considered implausible evolutionary adequacy is assessed negatively.

\item [Currencyness:] Does it qualify as a currency? This is only possible for an architecturally adequate candidate money which is evolutionary adequate to the extent that at some stage it constitutes a money. In that state exogenous events may switch the status of the informational money to being classified in addition as an informational currency. 

In Paragraph~\ref{PotentialC} below I will reflect on what it means for an informational money to be potentially an informational currency.
\end{description}

\subsection{Architectural adequacy  versus evolutionary adequacy}
Suppose that a system (design, architecture) $X_{mlic}$ constitutes a money-like informational commodity. And suppose we are interested in the question whether $X_{mlic}$ can evolve into a money. The there are two aspects to this: 
(i) the architectural potential of $X_{mlic}$ is determined by the degree to which in principle it might qualify as a money. This is a categorical judgement apriori that depends on one's theory of moneyness, 
and (ii) the ability of $X_{mlic}$ to survive and become established in an evolutionary process involving many existing and forthcoming competitors. The electric cars around 1910 were technically usable (mainstream and practical) cars but did not survive the evolutionary process in competition with combustion engine propelled  cars. Today's electric cars may be in different shape, but they are technically unproblematic, that is a car driving system where everyone uses today's electric car technology is conceivable from a point of technology.

Architectural potential moneyness may be denied to $X_{mlic}$ on many grounds. For instance in the case of Bitcoin someone may hold that the lack of authentication it the system undermines its very ability to qualify as a money under whatever conditions. Alternatively one may hold that the the level of security is too low, or that uncertainty regarding the origin of the system, which makes it very hard if not impossible to find out who has been responsible for critical design decisions about it, stands in the way of it being classified as a money. Obviously different individuals may have different opinions on this matter.

%
%
\subsection{Potential currency-ness: not an obvious concept}
\label{PotentialC}
The position of~\cite{Bergstra2014b} can be simply summarised as follows: 
for currencyness, that is the property of
constituting a currency, there is no distinction between the actual status and the potential status because it is
a matter of institutional formalization which either has or has not taken place.\footnote{%
This institutional formalisation is viewed as an event (state transition) which is exogenous to the built-in evolution of the system.}
And thus Bitcoin cannot be a currency-like informational commodity 
(that is a potential currency) without already qualify for being a currency (that is being an actual currency).

I am aware that my use of language is risky. At first sight an informational commodity is currency-like 
if it could become a currency in some plausible  future development. That is not what I have in mind, and the following comparison may be helpful for explaining what I have in mind instead.
A comparison may be made with a person $P$ supposed to be in the state $S_{A,t}$ of having performed an 
action $A$ at time $t$ in the past (that is being proven to have performed $A$ at historic time $t$). 
If $P$ is potentially in state 
$S_{A,t}$ then, even if $P$ is currently not in state $S_{A,t}$, $P$ may 
undergo a transition to state $S_{A,t}$ in the future with a non-zero probability. Assuming that assessment about past actions are made without false positives this state of affairs is only possible if in fact
person $P$ has performed action $A$ at time $t$.

Therefore person who potentially has performed an action in the past must have actually performed that action otherwise a transition to the state of having performed the action is impossible. The strength of this argument depends on the absence of false positives. That is in a state where ``$P$ has performed action $A$'', it must necessarily 
be the case that $P$ has performed $A$. Concerning the status of Bitcoin this means that the proposition that
``Bitcoin cannot be an currency-like informational commodity without already being a currency'' depends on the absence of false positives about the status of being a currency.  In spite of the logical complications involving a mix of probabilistic and temporal reasoning just mentioned, I trust that absence, as well as its role in the argument, sufficiently much to support the main thesis of~\cite{Bergstra2014b}.

\section{Ownability based classification of informational commodities}
Informational commodities may be classified in many different ways. Of course subject matter (what's the information about) and origin, accessibility and intellectual property all come into play. Below I will consider an attempt to classify  informational commodities in terms of the technical (authentication technology) and legal protection of the right to shared access and the right to exclusive access to the information making up quantities of the commodity at hand.

Many classifications of informational commodities are conceivable. I will speak of a quantity of an informational commodity when referring to a chunk of it which can be moved or traded or about which control can be lost or obtained. Quantities of a informational commodity are hosted by a, mostly distributed, underlying information system, in the case of Bitcoin that is a particular implementation of blockchain technology.

I will use the abbreviation QoIC for a quantity of an informational commodity. The particular commodity is apparent from the context.

\subsection{Authentication support and quantity ownability  as criteria}
I will propose a classification with respect to the strength of the binding to a legal entity. For that purpose I will distinguish the following three classes of informational commodities. For each class I will list a number of defining characteristics.
\begin{description}

\item [Exclusively informational commodity (EXIC):] an informational commodity is exclusively informational if there is no legal binding to any person possible and if there is no support for any binding of a specific quantity to any person or legal entity. Important characteristics of  an EXIC are these:\footnote{%
At this stage I will pay no attention to the question which of these characteristics, if any, might be considered defining characteristics.} 
\begin{itemize}
\item The EXIC comprises an underlying distributed information system (simply referred to below as the system), the users of which are involved in the maintenance and circulation of QoIC's, that is of control over QoIC's. 
\item Transactions within the system are about moving control around. Another way of looking at it is that QoIC's are moved through the system by means of transactions. 
\item All transactions are done by means of information transfer within the system.
\item All QoIC storage is in the form of information storage within the system.
\item Typically transactions go hand in hand with other transfers or transactions in the opposite direction rearguing control over or access to entities or services entirely outside the system. A QoIC is used in a money-like manner if its transfer can be considered a payment for a transfer in the opposite direction outside the system.
\item There is no ownership of any QoIC. The principal connection between agents and QoIC's is in terms of control of the agent over the quantity. 
\item An agent in control of a QoIC knows that it is in control of that QoIC. 
\item An agent may use its control over a QoIC to move it to a state where the agent probably has exclusive control over that same QoIC. Here it is assumed for the sake of simplicity  that some kind of identity may be assigned to the QoIC at hand which persists while its is moved though the system.
\item Control to a QoIC may be shared or exclusive. 
\item Agents have no guarantee on exclusive control over a QoIC. There is no test on exclusivity concerning an agent's control over a specific QoIC. 
\item Exclusive control (of agent $P$ over a QoIC $q$) comes close to ownership ($P$ owns $q$) but it is not the same and no-one is entitled to such exclusive control about any QoIC. 
\item Exclusive control by agent $P$ over a QoIC $q$ is called objective possession of $q$ by $P$.  If $P$ merely assumes that its control over $q$ is exclusive it has subjective possession over $q$.
\item If exclusivity (of $P$'s control over a QoIC $q$)  is lost due to whatever cause there is no claim possible (issued by $P$) against any perpetrator $Q$ who might be held responsible for that loss.
\item  In particular in case $Q$ after  gaining control over $q$ (against the wishes and intentions of $P$) moved $q$ outside control of $P,$ $P$ can't raise any legally backed claims against $Q$.
\item No authentication of any kind is incorporated in the underlying distributed information system. 
\item No legal support for any claims by (or against) holders of control over a quantity $q$ based on the fact that they have been (or still are) holders of $q$ is ever provided. 
\item Each change in the status of $q$  that can be achieved (by any agent say $Q$) by means of data processing is legalised in hindsight simply by it having been achieved. 
\item No action against any person is justified 
solely or even in part by whatever events took place  in the EXIC system.
\item No agent can be forced to perform an action within the EXIC.
\end{itemize}
The absence of authentication is not a guarantee of anonymity. I assume that within the EXIC's some systems are much better than other systems in guaranteeing the anonymity of system users.\footnote{%
The recent advance of negative interests in the Eurozone has brought about a renewed interest for cash as it
provides one with a way of storing money without the risk of being punished by negative interest rates. The anonymity of cash greatly contributes to that capability of it. It seems that within the class of EXIC-like MLIC's 
 the development of a further 
refinement of the classification w.r.t. the strength of anonymity is promising. In such a classification Bitcoin, when considered an EXIC (which is likely to be controversial) is likely to rank as less than optimal.}

\item [Strictly informational commodity (STRIC):] A strictly informational commodity differs from an EXIC only by allowing and possibly supporting information based authentication. Misleading that authentication is legally permitted (though it constitutes a misuse of the design objectives of the underlying system).
Defining characteristics of STRIC are these:
\begin{itemize}
\item Just as will an EXIC all transactions are done by means of information transfer within the system.
\item All QoIC storage is in the form of information storage within the system.
\item Control, objective possession (exclusive control), and subjective possession, for quantifies if a STRIC 
work just as for quantities of an EXIC.
\item Unwanted acquisition of control over a QoIC may be spotted and may be held against an agent (for instance through blacklisting), though not in a legally backed manner. A blacklisted agent may be refused as a trade partner for transactions related to the informational quantity.
\item No-one can take legal action against the system or any of its users based on having been a victim of impersonation or identity theft in the context of the system.
\item  Authentication may be required for a transaction but it will never exceed the digital boundary. That is no user will be physically searched or investigated in order to find out about the validity of authentication.
\item Wrong authentication is not held against the perpetrator in any manner. That is: a STRIC user may go at length to require authentication along a digital path but may not by legal means oppose any 
penalties on those who
have wrongly (but successfully) authenticated and acted on the basis of that authentication.

\item Users cannot be held accountable for whatever transactions they perform in the system. Promising to pay and not paying for instance cannot be punished in any manner, except the refusal to interact with the same person once more (modulo authentication).
\end{itemize}
For a QoIC of a STRIC, control over it and ownership of it cannot diverge for the simple reason that such QoIC's cannot be owned. However, to a greater extent than with an EXIC, the users of a STRIC may simulate the qualities of ownership on the basis of arbitrarily comprehensive authentication.

The relation between the concept of an EXIM and the concept of a STRIC is simple: 
an EXIM is a STRIC without authentication mechanisms.
\item [Ownable informational commodity (OWNIC):] each QoIC may or may not be 
in the state of being owned by a legal entity (person or otherwise). In case a QoIC is owned by an agent the ownership is protected by force of law. As a consequence there are notions of theft, 
damage, compensation, claim, etc. Some characteristic features of an OWNIC are these:
\begin{itemize}
\item Ownership of a QoIC is always exclusive (this is guaranteed by the underlying information system).

\item Ownership (of $q$ by $P$) implies control (of $P$ over $q$)  but may not imply exclusive control
(of $P$ over $q$). Ownership may have been compromised by 
unintended additional control that has been acquired by an adverse agent (say $Q$). Or alternatively an owner of a QoIC may have given additional control over it to a friendly agent $Q$.

\item An agent $Q$ who is having control over a non-owned QoIC in a manner not endorsed by its owner $P$  implies that $Q$ necessarily is in a forbidden state which allows (in principle, and upon detection of the fact) claims against $Q$ to be made and pursued. Such claims are also possible if the fact (illegal control of $Q?$ over some quantity $q$) occurred in the past.

\item Mere ownership must be distinguished from the preferred combination of ownership and possession.
\item The commodity may also allow occurrences of QoIC's  that are merely under control of 
but not owned by a legal entity.
That particular state of affairs  may occur as a consequence of theft. 
A stolen QoIC is not under the control of its owner.
\item Just as will an EXIC or with a STRIC all transactions are done by means of information transfer within the system.
\item All QoIC storage is in the form of information storage within the system.
\item Authentication of users is exchanged by informational means only, it may involve biometrical data and measurements.
\item There may be legally supported ownership on the basis of candidate 
owner authentication involving personal  inspection and biometrical data.
\item Users may exercise (legally validated) claims upon other users who don't operate as contracted, or who misuse the system. Exercising a claim of an agent $P$ against another agent $Q$ may involve applying some kind of force to a person or entity. That force may include such steps as to enforce one or more transactions in the OWNIC at hand, or with any other transactional system in part under control of $Q$ 
with the exclusion of EXIC's and STRIC's.

\end{itemize}

The description of an OWNIC includes the case that a QoIC is provided with a specification of an owner that
fails to apply to any existing agent or entity, in which case the quantity in fact has no owner or merely a fake owner. It may be imagined that the acquisition of access to the holdings of a fake owner is simpler than to the holdings of true owners. In the latter case (access to) a QoIC possessed by a fake owner  may plausibly be found by an agent in a way comparable to how a coin may be found which is has been dropped  on the street. 
Owner specification integrity is the property that an owner as indicated by the system corresponds to an existing agent or entity.

\end{description}

\subsection{What's the difference with gold?}
Assuming that gold constitutes a paradigmatic physical commodity class which may be thought of in connection 
with money, one may wonder where the main differences between a physical commodity and an informational commodity may lie.

Most importantly control over a quantity of a physical commodity (say a gold bar) is in most cases automatically, if not by definition, exclusive. Thus in the case of a gold bar control can be identified with possession, and any incentive to make use of the term control is absent. Possession and ownership must be distinguished, which is a classic distinction indeed. The major implication of the move towards informational commodities is that control 
replaces possession as a primitive category with control replacing possession in spirit and exclusive control serving as a definition of possession.

Similar considerations work if one replaces gold by units of physical money, coins, or banknotes alike.

\subsection{Further definitions}
Given the three classes of informational commodities one obtains a corresponding collection of informational moneys. In each case these are stages in the life-cycle of money-like instances of corresponding informational commodities. So one may have in mind the following notions:
  a money-like EXIC, 
 a money-like STRIC, and
 a money-like OWNIC.

Each of these may, once an MLIC has reached that stage, move through an evolution path where its moneyness fluctuates repeatedly from negative to positive  and back.
\begin{enumerate}
\item An informational money is an MLIC which  happens to be in a stage of its development where it constitutes an actual money. (In that stage the MLIC is obviously both architecturally adequate and evolutionary adequate for moneyness.)
\item An informational money is an exclusively informational money (EXIM) if and only if it is (based on) an 
exclusively informational commodity (EXIC).

\item An informational money is a strictly informational money (STRIM) if and only if it is an 
strictly informational commodity (STRIC).

\item An informational money is an ownable informational money (OWNIM) if and only if it is an 
ownable informational commodity (OWNIC).

\item  In~\cite{BergstraL2013}.
 the union of STRIM and OWNIM  is denoted with TIM (technically informational money).
\end{enumerate}

It is plausible that given an EXIM a variety of protocols can be implemented on top of it each of which provides a
subset of the EXIM's user community with a restricted and regulated view on the EXIM in such a manner that it
provides that part of the EXIM's user community with a STRIM.

\section{How about the status of Bitcoin?}
It is conceivable that both EXIM and STRIM will be prevented from coming into existence by authorities who feel that such systems are prone to fraud and tax avoidance. It is conceivable that only the prominence of an EXIM will be forbidden and that STRIM may come into existence while being disliked by people who feel that they need legal protection.
Currently I consider it plausible that on the long run (say 50 years) all money will develop into or be replaced by OWNIM.\footnote{%
This expectation does not take into account the effect of negative interests which makes anonymity of the
possession of quantities of money so much more important than it has been in the past.}

Assuming the moneyness of Bitcoin one may wish to classify it as an EXIM, a STRIM, or an OWNIM. As stated in~\cite{BergstraL2013} Bitcoin may be viewed as an EXIM if one so wishes. That classification, however, may have undesirable legal consequences which may lead to a preference for classifying Bitcoin as an OWNIM. 

For both categories STRIM and OWNIM it seems fair to say that Bitcoin is rather poor in terms of the way it facilitates the authentication of agents. Remarkably it is also problematic from the perspective of providing anonymity. It seems that Bitcoin obstructs authentication while not reliably providing anonymity. And by
combining these possibly unintended features Bitcoin seems to miss architectural adequacy.

It now seems to me plausible that in hindsight this particular combination of features (problematic authentication plus problematic anonymity) indicates design flaws in Bitcoin. I assume that the designers of Bitcoin have overestimated the anonymity that it will provide for its users and that it became only apparent via the practical use of a first generation of Bitcoin clients that anonymity has yet to be engineered into or on top of Bitcoin if it is to be fully reliable. If Bitcoin would provide guaranteed anonymity its classification as an EXIM would be justified to the point of being undeniable in my view.

\subsection{Assessment of current features of Bitcoin}
I believe that at this stage Bitcoin combines the following features:
\begin{enumerate}
\item  Bitcoin is an MLIC, 
\item Bitcoin is an EXIC-like MLIC,
\item some may classify Bitcoin as an EXIC but most people would prefer to view it as an OWNIC without authentication,
\item subjectively exclusive control of a QoIC that happens to be objectively exclusive as well  plays the role of its possession in Bitcoin,

\item Bitcoin is not a money (because of lack of circulation and acceptance), and therefore it is (at the moment of writing this paper)  not an informational money, not an EXIM, not an STRIM, and not an OWNIM.
\item Bitcoin is not a currency (because it is not a money),
\item it is (now) unknown if Bitcoin will  ever be classified as a money,
\item Bitcoin is potentially a money only if those in control of that concept will accept as a money either an 
EXIM (by definition without authentication support) or an OWNIM without authentication support. 
\item Bitcoin will need to compete with future designs for MLIC's in the design hull of Bitcoin
which provide either significant authentication mechanisms or reliable anonymity (as an optional feature) or both. Such variations on the blockchain technology theme are likely to outperform Bitcoin on shortly. The expectation of this forthcoming event indicates that even if Bitcoin is considered to be architecturally adequate is is not to be considered evolutionary adequate.
\item Bitcoin is probably not ever going to by a money, because (i) it is very unlikely that an EXIM will be allowed by regulating bodies while (ii) these will remain sufficiently strong to prevent the development to the state of 
moneyness of Bitcoin, and (iii) because any STRIM or OWNIM  will need to incorporate significant authentication mechanisms thus making it deviate from Bitcoin.
\item 
The use of an EXIM may be limited because some transactions, e.g. real estate transactions, may require a degree of authentication and proof which an EXIM or an EXIM-like MLIC cannot provide. If regulators accept an EXIM with limited functionality that may open the door to future moneyness of Bitcoin.
\item
Taken together the features of Bitcoin mentioned in the preceding items support its assessment as being  architecturally inadequate.
\end{enumerate}

\subsection{Expected eventual moneyness of Bitcoin and competitive evolution of MLIC's}
Based on a classification of informational commodities I have just motivated, somewhat to my own surprise,
 that I do not expect that Bitcoin will develop into a money because it is architecturally inadequate. At the same time I expect that 
within the design hull of Bitcoin new money-like informational commodities will be developed which will eventually achieve the status of money. 

That Bitcoin is architecturally inadequate is a bet on how it will be assessed in the future which can't be made with full confidence. Now in spite of this preliminary judgement of inadequacy one may introduce the working hypothesis that after all Bitcoin is architecturally adequate, which means that the appreciation of its features may evolve in such a manner that its adequacy will eventually be considered unproblematic.\footnote{%
In fact this means that to some extent architectural adequacy is an evolutionary concept too, though it is independent of the specific development of a particular candidate money. Architectural adequacy of a particular candidate money may fluctuate in line with fluctuating views on what constitutes a money.}

Whether or not Bitcoin will become money also depends on the alternative money-like informational commodities  that have already been developed  or that can be developed in the future, that is the present alt-coins as well as conceivable future alt-coins. Thus the question to predict future moneyness of Bitcoin leads to the question how to map the design hull of Bitcoin and to find out if it contains significantly stronger alternative designs which might be imagined and conceived, which are likely to be developed and which upon technical completion and distribution will eventually stand in the way of Bitcoin becoming a money by performing better in the competitive arena (market place) for monies.

In other words whether Bitcoin will reach the stage that it constitutes an informational money 
depends not only on Bitcoin per se but just as well on the development competing money oriented technologies.
I feel confident to guess  that money in the various forms of existence today 
will steadily develop in the direction of informational money. Bitcoin is definitely a remarkable step in that direction. But the situation may be similar to that with cars. Electric cars were used 100 years ago already,
but those models have become  outdated. Notwithstanding that a range of electric models has become obsolete it is stil very much possible that electric cars are essential for the future of sustainable automotive development.

\subsection{Architectural adequacy as a false positive yet useful working hypothesis}
An assessment that Bitcoin is architecturally inadequate constitutes  a bet on how it will be assessed in the future which can't be made with full confidence. Therefore in spite of the preliminary judgement of the architectural inadequacy of Bitcoin one may introduce the working hypothesis that, notwithstanding indications to the contrary, Bitcoin is architecturally adequate, which means that the appreciation of its features may evolve in such a manner that its architectural adequacy will eventually be considered unproblematic.\footnote{%
In fact this means that to some extent architectural adequacy is an evolutionary concept too, though when viewed as an evolving concept, it is independent of the specific development of a particular candidate money. Architectural adequacy of a particular candidate money may fluctuate in line with fluctuating views on what constitutes a money.}

In other words, providing a negative assessment on architectural adequacy for moneyness of an MLIC necessarily depends on
an underlying negative, or restrictive, assessment of the future flexibility of the concept of money. Even if one does not believe that the concept of money will prove to have some degree of flexibility there is the risk that a negative assessment on architectural adequacy is a false negative.

By making a positive assessment of the architectural adequacy for moneyness of Bitcoin (or any other MLIC), 
one runs the risk of producing working hypothesis which constitutes a false positive judgement. As a point of departure for searching the design hull of a given MLIC for stronger actual potential designs that false positive is harmless. Indeed once an actual or conceivable MLIC design has been spotted as a plausible future ``Bitcoin killer'', the plausibility of the latter assessment is only increased when taking into account  that the architectural adequacy of Bitcoin has been overestimated in an earlier phase of the search.

For everyone working with Bitcoin or doing research about Bitcoin the risk that its architectural inadequacy 
induces its termination constitutes a significant worry. But using a false positive working hypothesis concerning
Bitcoin's architectural adequacy as an incentive for searching the Bitcoin design hull for superior MLIC designs
to some extent neutralises precisely that risk. If Bitcoin survives, the importance of a systematic search for alternatives will only increase, while if Bitcoin fails to survive, a systematic search for alternative and superior designs will probably have  been even more meaningful and rewarding. 

\subsection{On predicting the future moneyness of Bitcoin; comparison with a different line of research}
Remarkably the question about the moneyness status of Bitcoin becomes significantly more interesting if one asserts that it currently isn't a money and that instead its potential for becoming a money must be assessed. Then first of all its
architectural adequacy must be assessed, a topic about which many remarks have been made in the literature already. Often economists claim that Bitcoin is based on a naive understanding of money, which to my
understanding can only signify that it is considered architecturally inadequate. Why exactly architectural inadequacy of Bitcoin can be claimed  is often not explained in any convincing detail, as if this were a matter that speaks for itself. 

Taking architectural adequacy (for moneyness) as  a working hypothesis in addition to the negative judgement about actual moneyness, the next phase of its classification 
is about making an assessment of its evolutionary adequacy.  In order to provide that particular kind of assessment  a survey is needed of the different options for developing informational moneys, that is a survey of the design hull around the Bitcoin design. If much better alternatives can be designed then it is reasonable to expect that such alternatives will be developed eventually and will eventually be victorious over Bitcoin as well.

This observation justifies an effort to investigate the design space of MLIC's
of which Bitcoin is merely a currently popular but otherwise somewhat ad hoc example. I consider
the challenge to contribute to that effort to be both attractive and intriguing. 

For some readers the search and survey project just mentioned 
may seem to be unbearably open-ended. Can one 
plausibly expect to find a useful structure theory for the design space of 
MLIC's which allows one to prove or to disprove that Bitcoin constitutes a meaningful local optimum in a relevant subspace of the design space?

However, this type of open-ended research question is definitely  attractive. Since 2007, I have been involved in an effort that may be described in similar terms. That effort merits being mentioned here extensively because it of its methodological correspondence with the effort that is required for an assessment of the evolutionary adequacy of Bitcoin in different and relevant ways. Working out this comparison will turn out to be useful in both directions.

In the next Section I will describe the project to which I propose to compare the assessment of the evolutionary adequacy of Bitcoin. Formulating that project in a  manner compatible with the terminology of the current paper requires some new formulations, however.\footnote{%
The comparison project deals with definitions of the concept of a fraction, and in particular with the evolutionary adequacy of a specific view on fractions the so-called 3D+ASA view, which is explained in detail below.

Comparing the evolutionary adequacy of Bitcoin with the evolutionary adequacy of some particular view of fractions may strike the reader as entirely unconvincing. Nevertheless both issues have some structure in common:
(i) independence of future development of technology, (ii) surveying the design hull can be done in both cases by means of theoretical work only  and it
involves no empirical research, (iii) and both themes feature an 
uneasy confusion between syntax and semantics. Concerning  the latter confusion the following may be noticed: (a)
mathematicians can't write about values without using expressions, though their theory is about values and not about expressions, (b)  In the world of Bitcoin there are no facts that aren't fully supported by a syntactic description. Still one (as a Bitcoin user) may feel an urge to say (or think): I have 100 BTC, I intend to use that QoIC for a payment, so where is my secret key? But there is no such state of affairs, because at the very moment the secret key may be missing the very ``having 100 BTC'' becomes implausible. Indeed one can hardly think about Bitcoin as a money without thinking in terms of access to values that may exist in the absence of access as well.

Finally although Bitcoin is competing with conventional monies the dominance of which is currently unchallenged, its survival is probably determined by how it will survive the competition with forthcoming design alternatives. I
support a corresponding perspective regarding the 3D+ASA view of fractions.}
It seems that casting the comparison project in terms of architectural adequacy and evolutionary adequacy in 
order to detail the comparison with the Bitcoin case is also helpful for acquiring a better understanding of the methodology of the comparison project, and what started out with an effort to provide an example meant for illustration from the perspective of informational monies, ends up with applying the methodological suggestions of this paper to a very different topic.

\section{Evolutionary adequacy of the 3D+ASA view of fractions}
Nowadays questions about the value of $1/0$ are largely ignored as irrelevant  by mathematics teaching staff. That attitude is by no means necessary or self-evident, however. A more general question is how to define the concept of a fraction in school mathematics. Only relative to a given a definition of fractions, one may expect progress regarding questions about the value, or values, of $1/0$. The question ``what is fraction'' seems to be easy but it is not. Fractions don't feature as a defined concept in most mathematical textbooks. 

I will briefly describe what constitutes to the best of my knowledge the currently dominant view on fractions. 
The dominant view at primary school level may be qualified as a 2D (2 dimensional) view, whereas the 
dominant view in later stages of mathematical maturity may be qualified as the value oriented, (or syntax avoiding) part of the so-called 3D view. Then I will describe a competing view, the ``3 dimensional view with alternating syntax appreciation'' (for short the 3D+ASA view) which has been proposed in~\cite{BergstraB2014} (though without use of dimension terminology). I will assume the architectural adequacy of
the 3D+ASA, and then the question becomes to assess the evolutionary adequacy of the 3D+ASA view   on fractions.

By assuming the architectural adequacy of the 3D+ASA view on fractions as a working hypothesis I assume
that the 3D+ASA view can be maintained as a coherent view on fractions. Assuming this working hypothesis goes at  the risk of making a false positive judgement. Now the situation is comparable to what I described before 
in the case of Bitcoin. If making a false positive judgement about the 3D+ASA view on fractions, phrased as a working hypothesis,  is used as an
incentive for searching the space of alternative views for superior ones, the risk of having wasted my time and energy upon a collapse of the working hypothesis by it being proven to be a false positive, is neutralised.
\subsection{Four views on fractions}
\label{ViewsF}
I will focus on fractions of non-negative natural numbers only, leaving out issues about subtraction and negative values. Then $p/q$ is a fraction with $p$ a natural number and $q$ a nonzero natural number. I will assume that
for a fraction $p/q$ the numerator is given by a function $\texttt{numerator}(-)$ defined by  
$\texttt{numerator}(p/q) = p$ and  the denominator is given by a function $\texttt{denominator}(-)$ with defining equation $\texttt{denominator}(p/q) = q$. Now, using $1/2 = 2/4$ one may obtain  the following chain of equations: $1 = \texttt{numerator}(1/2) = \texttt{numerator}(2/4) = 2$.

It is obvious that something is wrong with this chain of equations but it is less obvious what precisely  is wrong
with the argument that it contains.

My analysis of the case is as follows. In school mathematics a function $\texttt{numerator}(-)$ is not communicated as a component of the curriculum. For that reason there is no need to contemplate any of the consequences of its introduction. School mathematics 
is what I will call 2 dimensional (2D) by allowing the spacial form of texts as one dimension and the time in which
a student may transform the spatial forms as  the 2nd dimension. School mathematics, and in particular primary school mathematics is quite operational in nature and although it is stated that every fraction has a nominator the introduction of a function the extracts the nominator from a fraction is not included in the operations that a student is supposed to master.

If one accepts that each fraction has a numerator and a denominator and once one accepts that that very state of affairs implies the existence of extraction functions that decompose a fraction into such components then
the interpretation of $1 = \texttt{numerator}(1/2) = \texttt{numerator}(2/4) = 2$ requires further contemplation.
Now the following four views can be contrasted and compared:
\begin{description}
\item [Syntax appreciating view:] (also syntax oriented view, or expression oriented view) the fraction $p/q$ constitutes a pair (also called ``the expression $p/q$''), and such is the default reading of $p/q$.\footnote{%
It is left unexplained why some pairs are understood as fractions while other pairs are understood as a
point in a 2D plane, and yet other pairs are understood as points in the complex plane. In a richer context more typing is needed and viewing fractions as pairs suffices in a quite restricted context only.} The function $\texttt{numerator}(-)$ works on a pair (an not on a value of a pair view as a fraction) by extracting its first component. 

The equation $1/2 = 2/4$ is explicitly understood as asserting that the fractions $1/2$ and $2/4$ have equal values, not as  stating that both fractions are equal as fractions. Therefore the inference from $1/2 = 2/4$ to 
$\texttt{numerator}(1/2) = \texttt{numerator}(2/4)$  is mistaken because that inferences presupposes that $1/2$ and $2/4$ are equal as fractions.

By viewing a fraction as a pair the syntactic nature of $p/q$ is appreciated as a reading in support of say
the defining equation $\texttt{numerator}(p/q) = p$ for the numerator extraction function.
\item [Syntax avoiding view:] (also: value oriented view) the fraction $p/q$ is always understood as a rational value and the introduction of a function for extracting a numerator is rejected. Talk about numerator and denominator is merely explanatory and is not deemed worth of serious explanation. Therefore a chain of equations is not supposed to make use of functions like $\texttt{numerator}(-)$.

\item [Polymorphic view:] The expression $p/q$ is considered polymorphic as it can be attributed with different types. In the example at hand that is as follows. The equation $1/2 = 2/4$ casts (coerces) both the LHS (left hand side) and the RHS 
(right hand side) of the equation as values rather than as pairs. Therefore the inference from $1/2 = 2/4$ to
$\texttt{numerator}(1/2) = \texttt{numerator}(2/4)$ is unfounded as it confuses the two different type assignments 
of a fraction, as a pair and as a value. The context $\texttt{numerator}(1/2)$ casts (coerces) the expression $1/2$ as a pair. Type inference on the equation  $\texttt{numerator}(1/2) = \texttt{numerator}(2/4)$ reveals that it cannot be correctly typed and be valid at the same time.

\item [3D+ASA view:] The 3D view with alternating syntax appreciation. In this view a human observer regularly switches back and forth from the syntax appreciating view to the syntax avoiding view. Perhaps the polymorphic view is held in between during a phase of transition. The syntax avoiding view is dominant in terms of the length of the periods that it is held, which explains its role as default in mathematics. The 3D+ASA view oscillates between views that are inconsistent in combination. This phenomenon requires a dedicated logic able to avoid inconsistencies from blocking progress. Following~\cite{BergstraB2014} the chunk and permeate reasoning pattern for paraconsistency from~\cite{BrownP2004} provides precisely that kind of approach to logic.

The 2D/3D terminology arose from a comparison with the famous Necker cube due to~\cite{Necker1832} (see
also~\cite{Mortensen2006}). That is a 2D graph which invites a human observer to maintaining alternating 3D interpretations of it. The 2D operational view of calculating with fractions invites one to have a deeper understanding, which unavoidably results in alternating views switching between the syntax appreciating view and the syntax avoiding view.\footnote{%
Stated differently: a view regularly switching from the expresion 
oriented view to the value oriented view and back.}

\end{description}

\subsection{Comparing  3D+ASA evolutionary adequacy with that of Bitcoin}
So I wish to put forward the following comparison: (i) the syntax avoiding view as the mainstream conventional view on fractions is compared with conventional money, (ii) the 3D+ASA view is a contender for prominence 
in the ``arena'' of views on fractions, (iii) the hypothesis that Bitcoin is architecturally adequate for moneyness
is a useful basis for investigating its evolutionary adequacy, just as the hypothesis that the 3D+ASA view is architecturally adequate (for constituting a coherent view on fractions) provides a basis for investigating the evolutionary adequacy of the 3D+ASA view, (iv) the syntax appreciating view and the polymorphic view are alternative design options for a non-conventional view on fractions, (v) searching the space of design options for a view on fractions will be informative for estimating the evolutionary adequacy of  the 3D+ASA view, and (vi) the polymorphic view is a potentially stronger 
contender of the 3D+ASA view in much the same way as a well-designed STRIM may constitute a potentially stronger contender of Bitcoin.

The comparison that I am making is Bitcoin as a plausible MLIC with the 3D+ASA view on fractions. Rather than to formulate the research question to find a better alternative to the dominant view on fractions I prefer to formulate the question about its long term sustainability and to take that latter question as an incentive to perform an in depth survey of the design alternatives for the 3D+ASA view with the expectation that something novel and useful can be found during that search.

\subsection{Implications for division by zero}
I hold that most mathematical courses deal with the phenomenon of division by zero in a purely semantic way.
That is an expression like $1/0$ is not entitled to any attention because there is no meaning, that is semantic part of the world, which apriori ``asks'' (invites one) for being denoted that way. That $1/0$ constitutes an expression which for that reason justifies, or even necessitates, an investigation of its possible meanings is not seen as remotely convincing. This view on $1/0$, that is on division by zero, is compatible with the the syntax avoiding view on fractions. That view on $1/0$ may be called the syntax avoiding view on division by zero.

As an alternative to the syntax avoiding view one may consider the syntax appreciating view on division by zero. In the syntax appreciating view $1/0$ is considered to be an expression and contemplating the meaning of that expressions makes very much sense.
Now the question may be asked: is the syntax avoiding view on division by zero sustainable on the long run, 
or will it be replaced by a different and syntax appreciating view. 

This question may be understood as a forecasting problem, for which a preparatory analysis of the evolutionary adequacy of the 3D+ASA view on fractions will be helpful.

I like this question because I hypothesise that the answer wil be negative, that is the syntax avoiding view will be replaced by a syntax appreciating view. This will happen under the pressure of informatics which will slowly but steadily replace logic as the foundational basis for school mathematics. The difficulty with focusing on $1/0$, however, which makes it useful to think in terms of views on fractions instead, is that each of the various design alternatives for a theory on the value of $1/0$ seem to be overly specific when compared with the conventional view, a phenomenon that seems not to arise in the case of that variety of views on fractions.

In order to make progress on the viability of the syntax avoiding view on division by zero, the development of  a survey of alternative designs of a story about fractions including the respective  implications on division by zero is needed. Determination and subsequent analysis of the different explanations of division by zero in the design hull of the 3D+ASA view on fractions is probably simpler than determination and subsequent analysis of the architectures in the design hull of Bitcoin. But in the case of division by zero (and fractions) it seems to be a manageable effort and which in addition proves to be an intriguing and unpredictable excursion into unexpected options. 

The following papers are among the results of my own contribution to that search and much more has been done and can be done:~\cite{BergstraT2007,BergstraM2015,BergstraP2014, BergstraB2014,BergstraP2014b}. Other results include~\cite{Ono1983,Carlstroem2004,Carlstroem2005,ReisA2014,Setzer1997}. Slowly but steadily all alternative views to division by zero are brought into perspective and a condition is reached from which the question about the sustainability of the syntax avoiding view can be discussed with some confidence.

Now the topic of design alternatives for stories concerning division by zero is all but exhausted with the references just mentioned. In spite of the low number of papers just mentioned, acquiring a workable survey of the complete space of design alternatives in this case, to such an extent that no useful options are being overlooked, is not at all easy. 

\subsection{Implications for a theory of fractions}
Remarkably, by proposing the issue of assessment of the evolutionary adequacy of the 3D+ASA view on fractions as an issue which can be productively compared with the issue of the assessment of the evolutionary adequacy of Bitcoin, the question about fractions is placed in a novel perspective. That perspective does away with an oversimplification I apparently had in mind when favourably contemplating the 3D+ASA view, an oversimplification which may perhaps be compared to my initial enthusiasm about Bitcoin. To begin with more alternative designs of a story on fractions need to be looked for and contemplated in addition to the one's mentioned above. Here are some options:
\begin{description}
\item [Expression-value ambiguity:] expression and value are two distinct readings of say $1/2$. This particular ambiguity must be taken into account with some care but it does not stand in the way of any proper understanding. This is a specific case of ambiguity (unlike say light in connection with sunlight and light in connection with weight) because the contrasting pair of meanings from which the ambiguity arises (as an expression and as a value) are intrinsically related.
\item [Merely a case of ambiguity:] ambiguity of meaning occurs  with many words and one needs to deal with it in a general and uniform manner. That general (generic) approach applies just as well in this case. There is no need to use or develop a specific theory in view of the interconnection of ``expression $1/2$'' and ``value $1/2$''
\item [Process-product ambiguity w.r.t. fracturing:] The value $1/2$ may be considered the product of the process of fracturing an entity with size $1$. The expression $1/2$ may be understood as a snapshot (one among many) of the process of fracturing.
\item [Process-product ambiguity w.r.t writing:] The expresion $1/2$ may be viewed as a product of writing (about a simple fraction), while the value $1/2$ may be understood as an aspect of the process that goes along with the mental processes behind writing $1/2$ (when understood as a human action).
\item [Process-product ambiguity w.r.t. imagination:] one may interchange product and process in connection with fractions, viewing the value $1/2$ as the product of meaning assignment and the expression $1/2$ as a a step in the process towards the generation of that product.
\item [The 2D view of fractions as embedded network elements:] one may abandon the idea that $1/2$ is a 
textual fragment entitled to
its own syntactic and semantic analysis. A text fragment like $1/2$ can only occur as embedded in a larger context, say a graph, of text fragments. Just as a theory of fractions need not contemplate ``subexpressions'' like
$1/$ and $/2$ a theory of embedded fractions need not pay much attention to non-embedded fractions. In this manner a reconciliation with the operational 2D view can be found.
\end{description}

Investigating each of these options, in addition to the alternative views that were mentioned in~\ref{ViewsF}
above, is needed if an assessment of the future sustainability of the 3D+ASA view is to be developed.

The architectural adequacy of the 3D+ASA view cannot be taken for granted. Indeed if the comparison with the Necker cube is fully valid one must expect that the alternation of syntax appreciation and syntax avoidance 
have some neurological basis and can be demonstrated by means of neuroimaging technology. Conversely the
architectural inadequacy of the 3D+ASA view may be disproved if a neural correlate turns out to be non-existent.  

\subsection{Carrying on with fractions, and back to informational monies}
The idea that a valid view on fractions is found by choosing the optimal view, i.e. most evolutionary adequate view, from a sample of  architecturally adequate views, which in turn have been selected from a larger collection
of candidate views, may be questioned. It may just as well be the case that a human agent maintains a weighted combination, perhaps with dynamically adaptable weights, of a range of different architecturally adequate views on fractions. Perhaps even some architecturally inadequate views survive within a human mind with non-zero weight factors. 
On needs a (meta)view on views on fractions in order to make progress at this level of abstraction.

The same is true for informational moneys. The above discussion of Bitcoin has as a hidden assumption that Bitcoin, or any MLIC can't survive as a minor niche player on the long run. But that is not necessarily true, perhaps Bitcoin will become important in between of thousands of equally important competitors. Without a perspective on the ecology of multi-MLIC systems it may not be possible to arrive at a convincing assessment of 
its evolutionary adequacy. Can it be the case that a weakness of Bitcoin gives rise to the emergence of a competitor with which Bitcoin co-evolves into a mutually beneficial and sustainable co-existence? Can it be the case that a feature that seems to constitute a weakness of Bitcoin in fact constitutes a major competitive advantage through a detour involving the co-evolution of another informational money which subsequently constitutes an invincible tandem with Bitcoin? 

\subsection{Infinite regress: with views on fractions, and with MLIC design}
These questions suggest that the scenery of existing and forthcoming informational monies is far more complex
than the scenery of views on fractions. However, that may not be true as it is easy to see that the perspective on views on fractions that has been sketched in the previous Paragraphs  is too simple.

An experimental calculus of views on fractions may be introduced as follows. $\FR$ denotes the as yet unanalyzed concept of a fraction, an approximation of which is given in~\cite{Padberg2012} with the labeling of fraction as a complex concept that can be viewed from a range of different perspectives. Then $EO_v(\FR)$ may denote the expression oriented view on $\FR$ and $V\!O_v(\FR)$ may denote the value oriented view on $\FR$.
With $(3D+ASA)_v(\FR)$ one may denote the 3D+ASA view on fractions, which involves a dynamic alternation
of  $EO_v(\FR)$ and $V\!O_v(\FR)$.

It seems undeniable that $EO_v(\FR)$ constitutes a component of $(3D+ASA)_v(\FR)$ which itself must 
be an architecturally adequate view on fractions, though not necessarily a potentially dominant one. In $EO_v(\FR)$
a fraction is a typed (as a fraction) pair $(p,q)$ of expressions $p$ and $q$ for integers (conventionally under 
the assumption that the value of $q$ is non-zero.) . 

At this stage one may notice that it is attractive to view $(p,q)$ simply as a pair of integers, and not to be bothered by a syntactic view on those parameters. This perspective leads to  a value oriented view on $EO_v(\FR)$,
which may be denoted as  $V\!O_v(EO_v(\FR))$. If pairs of integers serves as values for expressions for fractions, what are expressions for expressions for fractions? One may distinguish $\frac{p}{q}$, $p/q$, and $p:q$ 
as different expressions (for expressions) for fractions and then one may introduce a third extractions function
$\texttt{op-symb}(-)$ which produces a code for the operator symbol. Codes are $\{\texttt{hb,sl,co}\}$, respectively indicating horizontal bar, slash, and colon. Now using $V\!O_v(EO_v(\FR))$ one finds $1/2 = 1:2$, and one may derive $\texttt{sl} = \texttt{op-symb}(1/2) = \texttt{op-symb}(1:2) = \texttt{co}$. This chain of equations must be wrong and
the apparent reasoning anomaly is very similar to the reasoning anomaly noticed in~\ref{ViewsF} above.

Just as  with the corresponding anomaly involving the numerator function one may propose a view that alternates between 
$V\!O_v(EO_v(\FR))$, and $EO_v(EO_v(\FR))$. The latter view may be denoted with 
$(3D+ASA)_v(EO_v(\FR))$.\footnote{%
It now appears that more precision with typing is useful. By writing 
$(3D+ASA)_v(EO_v(\FR), V\!O_v(\FR))$ instead of $(3D+ASA)_v(\FR)$ it is made explicit that a recurring alternation between the view $EO_v(\FR)$ and the view $ V\!O_v(\FR)$ is meant. Taking operator symbols into account as a constituent of expressions (for expressions of type $\FR$), one arrives at a view that may be denoted with
$(3D+ASA)_v((3D+ASA)_v(EO_v(\FR)), V\!O_v(\FR))$ or more consistently with 
$(3D+ASA)_v((3D+ASA)_v(EO_v(EO_v(\FR)), V\!O_v(EO_v(\FR)),V\!O_v(\FR))$. This representation of the nested view highlights that  paraconsistent reasoning patterns appear in a nested and layered manner. 
In turn these expressions for views may be targets for contrasting a value oriented view and an expression oriented view, thereby sprouting the infinite regress of view refinement in a further dimension.}

At this stage we are looking at the beginning of an infinite regress. Indeed $EO_v(EO_v(\FR))$ may be complemented (enriched) with a view where an expression has a size that measures either the length of the horizontal bar, or the length of a slash, or the distance of both parts of the colon. With this additional level of detail one may wish to distinguish views $EO_v(EO_v(EO_v(\FR)))$ and $V\!O_v(EO_v(EO_v((\FR)))$ an so on.

The implication of this observation seems to be that, at closer inspection, 
it may turn out to be impossible to formulate a coherent (that is 
architecturally adequate) and purely expression oriented view on fractions.\footnote{%
This seems to be different with value oriented views on fractions. A classroom tested and purely value oriented view on fractions is proposed in~\cite{Rollnik2009}.} Unless an infinite regress is accepted at some layer of depth of conceptual decomposition a value oriented view must be accepted. The infinite regress need not be prohibitive if decreasing weight factors are assumed. But the
implications of these observations (when valid) are nevertheless remarkable. Unless one accepts a value oriented view at some level, which will unavoidable give rise to ``paradoxes'' of the kind that leads to the introduction of alternating views at that stage, one is led to maintain a deeply nested cascade of alternating views each of which is justified by its own specific regime of paraconsistent reasoning.

Going back to informational monies one may ask if infinite regress phenomena feature in that setting as well? That seems to be the case as soon as one proposes to measure the development cost of an informational money in terms of a previous informational money. Can it be the case that sophisticated informational monies must be developed in order to measure and distribute the cost of the development of even more sophisticated informational monies? Can it be the case that in this manner Bitcoin constitutes a necessary stage of development which facilitates further development and evolution of other MLIC's in a manner that could not have been provided with any conventional money?

\section{Concluding remarks}
The situation with Bitcoin and an attempt to survey its design alternatives in its quality of a money-like informational commodity is much more complicated than the case of fractions and division by zero. 
But in essence both forecasting issues are somehow comparable. Both issues have in common that the relevant  design space is not 
defined in terms of technological options but in terms of conceptual options. And that implies that a search can be performed in principle with limited means and by making use of today's view of underlying technologies only.

Further and at least as important, both in the case of Bitcoin and in the case of the 3D+ASA view of fractions the line of though that has been proposed centring around architectural adequacy and evolutionary adequacy 
provides a methodology for investigating both views (the Bitcoin view of informational money and the 3D+ASA view of fractions) in a manner that does not force one to be overly committed to either view.

\bibliographystyle{plain}

\end{document}